\documentclass[lettersize,journal]{IEEEtran}
\usepackage{amsmath,amsfonts,amssymb}
\usepackage{algorithmic}
\usepackage{algorithm}
\usepackage{array}
\usepackage[caption=false,font=normalsize,labelfont=sf,textfont=sf]{subfig}
\usepackage{textcomp}
\usepackage{stfloats}
\usepackage{url}
\usepackage{verbatim}
\usepackage{graphicx}
\usepackage{cite}
\usepackage[T1]{fontenc}
\usepackage{graphicx}
\usepackage{multirow}
\usepackage{mathrsfs}
\usepackage{dsfont}
\usepackage{makecell}
\usepackage{paralist}
\usepackage{todonotes}
\usepackage{makecell}
\usepackage{tikz,xcolor}
\usepackage[hidelinks,
            colorlinks=False,
            linkcolor=black,       
            anchorcolor=black,  
            citecolor=black,       
            ]{hyperref}
\usepackage{booktabs}
\definecolor{lime}{HTML}{A6CE39}
\DeclareRobustCommand{\orcidicon}{%
	\begin{tikzpicture}
	\draw[lime, fill=lime] (0,0) 
	circle [radius=0.16] 
	node[white] {{\fontfamily{qag}\selectfont \tiny ID}};
	\draw[white, fill=white] (-0.0625,0.095) 
	circle [radius=0.007];
	\end{tikzpicture}
	\hspace{-2mm}
}

\foreach \x in {A, ..., Z}{%
	\expandafter\xdef\csname orcid\x\endcsname{\noexpand\href{https://orcid.org/\csname orcidauthor\x\endcsname}{\noexpand\orcidicon}}
}

\begin{document}

\title{DVG-Diffusion: Dual-View Guided Diffusion Model for CT Reconstruction from X-Rays}

\author{Xing Xie$^*$\orcidA{}, %\IEEEmembership{Fellow, IEEE},
Jiawei Liu$^*$\orcidB{}, 
Huijie Fan\orcidC{}, %\IEEEmembership{Member, IEEE}
Zhi Han\orcidD{}, 
Yandong Tang\orcidE{}, 
and Liangqiong Qu\orcidF{}

\thanks{This work is supported by the National Natural Science Foundation
    of China (62306253, U23A20343, U20A20200, 61873259), Guangdong Natural Science Fund-General Programme
(2024A1515010233), and the Youth Innovation Promotion
    Association of Chinese Academy of Sciences(Y202051). \textit{(Xing Xie and Jiawei Liu contributed equally to this work.) (Corresponding authors: Liangqiong Qu; Zhi Han)}}
% \thanks{Xing Xie and Jiawei Liu contributed equally to this work.}
% \thanks{Liangqiong Qu and Zhi Han are the corresponding authors.}
\thanks{X. Xie and J. Liu are with State Key Laboratory of Robotics,
    Shenyang Institute of Automation, Chinese Academy of Sciences, Shenyang,
    110016, China;
    University of Chinese Academy of Sciences, Beijing, 100049, China
    (email: (xiexing,liujiawei)@sia.cn).}
\thanks{H. Fan, Z. Han and Y. Tang are with State Key Laboratory of Robotics,
    Shenyang Institute of Automation, Chinese Academy of Sciences, Shenyang,
    110016, China; (email: \{fanhuijie,hanzhi,ytang\}@sia.cn).}
\thanks{L. Qu is with the Department of Statistics and Actuarial Science and
    the Institute of Data Science, The University of Hong Kong, Hong Kong,
    999077 (e-mail: liangqqu@hku.hk).}}

\markboth{Journal of \LaTeX\ Class Files,~Vol.~14, No.~8, August~2021}%
{Shell \MakeLowercase{\textit{et al.}}: A Sample Article Using IEEEtran.cls for IEEE Journals}


\maketitle

\begin{abstract}
Directly reconstructing 3D CT volume from few-view 2D X-rays using an end-to-end deep learning network is a challenging task, as X-ray images are merely projection views of the 3D CT volume. In this work, we facilitate complex 2D X-ray image to 3D CT mapping by incorporating new view synthesis, and reduce the learning difficulty through view-guided feature alignment. Specifically, we propose a dual-view guided diffusion model (DVG-Diffusion), which couples a real input X-ray view and a synthesized new X-ray view to jointly guide CT reconstruction. First, a novel view parameter-guided encoder captures features from X-rays that are spatially aligned with CT. Next, we concatenate the extracted dual-view features as conditions for the latent diffusion model to learn and refine the CT latent representation. Finally, the CT latent representation is decoded into a CT volume in pixel space. By incorporating view parameter guided encoding and dual-view guided CT reconstruction, our DVG-Diffusion can achieve an effective balance between high fidelity and perceptual quality for CT reconstruction. Experimental results demonstrate our method outperforms state-of-the-art methods. Based on experiments, the comprehensive analysis and discussions for views and reconstruction are also presented.
\end{abstract}

\begin{IEEEkeywords}
    Diffusion model, CT Reconstruction, X-rays, 2D-to-3D
\end{IEEEkeywords}

\section{Introduction}
\IEEEPARstart{C}{omputed} tomography (CT) is a widely used imaging technique in medical diagnosis, and can provide detailed internal images of various body parts \cite{sluimer2006computer}.
However, the scanning process of CT exposes patients to large amounts of ionizing radiation \cite{li2020visualization}, and the high cost of CT scanners also limits their widespread availability in some clinics and developing countries.
Additionally, CT requires patients to be scanned in a lying position, which is unsuitable for some special patients and diseases, such as spinal disorders \cite{chen2023bx2s}.
In contrast, X-ray imaging is one of the most common examinations in hospitals, especially in clinics, offering  advantages such as low radiation, affordability and wide availability for patients.
However, X-ray imaging provides only limited 2D information without 3D view, which limits its availability for clinical diagnosis, especially for early diagnosis of diseases.

In recent years, AI-driven medical image synthesis has achieved remarkable successes~\cite{feng2023learning,guo2021multi,jun2021joint,qu2019wavelet,kong2021breaking,qu2020synthesized,hait2018spectral}, including CT reconstruction from a single-planar or biplanar X-rays~\cite{jiang2022mfct,kyung2023perspective,ying2019x2ct,ratul2021ccx,shen2019patient,corona2022mednerf,ge2022x}. Ying \emph{et al}. \cite{ying2019x2ct} pioneered the X2CT-GAN, demonstrating that generative adversarial network (GAN) can be used to learn the X-rays-to-CT mapping in an end-to-end manner. Since then, various subsequent techniques have been further developed to improve CT reconstruction, such as incorporating depth information~\cite{ratul2021ccx}, multitask learning \cite{ge2022x}, and shape estimation schemes~\cite{jecklin2022x23d}. These methods enhance reconstruction performance with estimated depth, segmentation map, or other information.
% For example, Ratul \emph{et al}. \cite{ratul2021ccx} utilized additional organ segmentation maps extracted from X-ray images to provide depth information for enhanced CT reconstruction. 
However, they rely heavily on the accuracy of the estimated information, which leads to performance degradation when the acquired prior is inaccurate, affecting their applications.

To avoid reliance on estimation information, some researchers attempt to directly learn the mapping between 2D X-ray images and 3D CT volume. They designed multichannel residual block \cite{jiang2022mfct}, feature attention module \cite{chen2023bx2s} and transformer module \cite{wang2023trct}, achieving promising reconstruction results.
However, these direct X-rays-to-CT mapping methods are inherently challenged by the lack of one-to-one correspondence, as the X-ray image is only a projection view of the 3D CT volume, which compromises reconstruction accuracy. %lacks one-to-one correspondence, as the X-ray image is only a projection view of the 3D CT volume, affecting the accuracy of the direct mapping. 
In \cite{ying2019x2ct,ratul2021ccx}, Ying \emph{et al}. highlighted that providing more real X-ray views as input can significantly enhance CT reconstruction, as these extra views offer complementary information to reduce the learning ambiguity. Despite these findings, acquiring multiple X-ray views for a single patient is often impractical due to constraints such as increased time costs, patient discomfort, and the need for efficient medical workflows. This practical barrier prompts a critical, yet underexplored, question: \textit{Instead of relying solely on the limited available X-ray
views for CT reconstruction, can the synthesis of additional X-ray views and their joint utilization with the original input view facilitate
the CT reconstruction process?} % and the offset errors during imaging affect actual reconstruction precision.

To answer the above question, we propose a novel approach called the Dual View Guided Diffusion model (DVG-Diffusion), which couples real input X-ray views with a synthesized new X-ray view to jointly guide CT reconstruction and refinement. 
%Our model couples real input X-ray views with a synthesized new X-ray view to jointly guide CT reconstruction and refinement. 
The core of our DVG-Diffusion lies in a novel View Parameter Guided Encoder (VPGE) that back-projects the X-rays into a 3D latent space aligned as closely as possible with the CT latent space. As shown in Fig. \ref{fig:x2ct framework}, in DVG-Diffusion, a new X-ray view is first synthesized from the given X-rays (Fig. \ref{fig:x2ct framework}(b)). The VPGE module generates dual-view features by extracting features from both the synthetic X-ray view and the input X-ray view. These dual-view features serve as conditions for the latent diffusion model to learn and refine the CT latent representation, which is then decoded back into pixel space for CT reconstruction. Overall, we utilize the additional synthesized X-ray view and leverage the VPGE module to establish spatially coherent representations between X-rays and CT modalities. Generation in aligned latent space ensures the perceptual quality, while additional view guidance enhances the fidelity. By this way, our DVG-Diffusion model achieves a remarkable balance between high fidelity and high perceptual quality for CT reconstruction.

The main contributions are summarized below: 
% (\romannumeral1) We propose a novel dual-view guided diffusion model (DVG-Diffusion), marking the first instance of integrating the additional view of synthesized new X-rays into CT reconstruction, aiming to provide more complementary information.

\begin{itemize}
    \item We propose a novel dual-view guided diffusion model (DVG-Diffusion), marking the first instance of intergrating the additional view of synthesized new X-rays into CT reconstruction. This synthetic view provides complementary information, thereby enhancing the reconstruction performance.
    \item We propose a view parameter guided encoder (VPGE) that back-projects the X-ray into a 3D latent space that aligns more readily with the CT latent space. This approach transforms the original intricate 2D-to-3D pixel space mapping into a more feasible 3D-to-3D latent space mapping.
    \item Experimental results show that our method achieves state-of-the-art results, realizing the effective reconstruction of the 3D structure. Our experiments also provide valuable insights about the relationship between views and reconstruction.
\end{itemize}

\section{Related Work}
The conversion from 2D images to 3D volumes is a challenging task, as it involves recovering volumetric information from a limited number of projections. We review the methods for 3D reconstruction in CT and natural images, respectively.
\subsection{3D CT Reconstruction from 2D X-Rays}

Currently, the most promising solution for learning X-ray-to-CT mappings is deep learning. Initial work by Henzler \emph{et al}. \cite{henzler2018single} proposed a 2D deep neural network (2DCNN) for reconstructing 3D volumes from 2D X-ray images. %, but did not apply it to clinical diagnostic CT data of real patients. 
Shen \emph{et al}. \cite{shen2019patient} improved 2DCNN by using patient-specific CT images for training and introducing a dimensional change module to facilitate 2D-3D mapping. 
However, these CNN-based methods rely on pixel-level constraints, potentially leading to overly smooth CT images. To address this issue, Ying \emph{et al}. \cite{ying2019x2ct} employed adversarial loss in their X2CT-GAN model, enhancing the perceptual details of reconstructed CT images. Since then, several other GAN-based models were developed \cite{ratul2021ccx, jiang2022mfct,wang2023trct}. However, the samples generated by GAN-based models tend to focus on perceptual aspects rather than the fidelity of the reconstructed images, such as the essential structural information that is critical for clinical diagnosis.
%Therefore, Ying \emph{et al}. \cite{ying2019x2ct} used adversarial loss in their X2CT-GAN model to enhance the perceptual details of reconstructed CT. Several other GAN-based models have since been developed \cite{ratul2021ccx, jiang2022mfct,wang2023trct}. However, the samples generated by GAN models tend to be too perceptual and ignore the real structural information, which is fatal for clinical diagnosis.

To enhance fidelity, some works applied estimation information, such as depth maps and segmentation maps, to guide learning \cite{ratul2021ccx,jecklin2022x23d,ge2022x}. Ratul \emph{et al}. \cite{ratul2021ccx} used additional organ segmentation maps to provide depth information for improved CT reconstruction. 
% Jecklin \emph{et al}. \cite{jecklin2022x23d} introduced a new shape estimation method, X23D, which uses multi-view sparse X-ray data and corresponding image calibration parameters as input to extract 3D shape representation of the lumbar spine. 
Ge \emph{et al}. \cite{ge2022x} designed a set of multi-task learning strategies to enhance CT reconstruction by coupling the reconstruction network with the segmentation network. These methods aim to learn the X-ray image to CT mapping in an end-to-end manner, guided by estimated information.
% However, these prior are not easily accessible in clinical diagnostics, limiting the applicability of such methods.
However, the reliance on the accuracy of estimation limits their clinical values.

Mainstream GAN-based methods suffer from low fidelity, while CNN-based methods do not ensure optimal perceptual quality. In this paper, we introduce a new perspective to the challenging 2D-X-ray-to-3D-CT task by decomposing it into simpler sub-tasks, thereby significantly reducing learning difficulty and achieving a balance between fidelity and perceptual quality. Our novel task decomposition involves synthesizing new views using readily available view parameters in clinical diagnostics, translating intricate 2D-to-3D mapping into a more feasible 3D-to-3D latent space mapping, and then reconstructing and refining CT volumes. We anticipate that this strategy of task decomposition could provide a new learning paradigm for CT reconstruction field.
%In addition, we integrate view parameters that is easily accessible in clinical diagnostics into the model, and exploited potential information through new-views.  Our approach provides a new learning paradigm for this field.

\subsection{Few-view 3D Reconstruction in Natural Images}
% The computer vision community has made exponential progress in generative modelling. Various methods based on GAN, VAE, Flow, and autoregressive models have been proposed to achieve high quality image generation. Recently, diffusion models have obtained state-of-the-art results in several image generation tasks. The denoising process of diffusion models obtains better distribution estimates while avoiding pattern collapse. Our goal is to extend this powerful image diffusion model from 2D image synthesis to 3D content generation and inference.
Similar to X-ray-to-CT task, few-view 3D reconstruction task in natural images aims to recover the 3D structural from a limited number of 2D images. 
Due to the limited view information, traditional 3D reconstruction methods using polar constraints \cite{hartley2003multiple} cannot accurately determine spatial information.
Recently, diffusion models \cite{ho2020denoising} have achieved state-of-the-art results in several image generation tasks \cite{bar2024lumiere,gao2024multi, zhang2023inversion, ding2023diffusionrig,chan2024anlightendiff,wang2025diffusion}, including 3D reconstruction from few-view 2D images \cite{anciukevivcius2023renderdiffusion, qian2023magic123, liu2023deceptive, ahn2022panerf, gu2023nerfdiff, poole2022dreamfusion,metzer2023latent}. Some classical approaches typically use NeRF as the 3D representation and 2D diffusion model as a prior for reconstruction \cite{gu2023nerfdiff,ahn2022panerf,liu2023deceptive}. However, these approaches focus on reconstructing individual scenes and lack generalization to other images and scenes. Some works utilized additional learning networks to achieve generalized generation \cite{yu2021pixelnerf, yao2018mvsnet}.

% Multi-view 3D reconstruction aims to recover the 3D structure of a scene from its 2D RGB images captured from different camera positions \cite{faugeras1992can,agarwal2011building}.
% Recently, the computer vision community has made exponential progress in few view 3D reconstruction through the generative model (GAN, Diffusion model).

% Classical approaches usually recover a scene’s geometry as a point cloud using SIFT-based \cite{lowe2004distinctive} point matching\cite{schonberger2016pixelwise,schonberger2016structure}.
% More recent methods enhance them by relying on neural networks for feature extraction (e.g. \cite{yao2018mvsnet,yao2019recurrent,yu2020fast,huang2018deepmvs}). The development of Neural Radiance Fields (NeRF) \cite{mildenhall2021nerf,lombardi2019neural} has prompted a shift towards reconstructing 3D as volume radiance, enabling the synthesis of photo-realistic novel views. Subsequent works have also explored the optimization of NeRF in few-shot (e.g. \cite{du2023learning,kim2022infonerf,jain2021putting}) and one-shot settings. NeRF does not store any 3D geometry explicitly (only the density field), and several works propose to use a signed distance function to recover a scene’s surface, including in the few-shot setting as well (e.g. \cite{yu2022monosdf,zhang2021ners}).

Although diffusion models in natural images can be optimized for NeRF to achieve few-view 3D reconstruction, the NeRF-based approach is not directly applicable to CT reconstruction from X-rays. This is due to the penetration of X-rays through most objects, which results in the superimposition of different structures onto a 2D image. Consequently, the methods for extracting 3D models from X-ray images differ significantly from those employed in natural images. Additionally, NeRF-based implicit reconstruction methods can only capture the external shape and are unable to obtain the internal 3D details required for CT reconstruction.
Corona \emph{et al} \cite{corona2022mednerf} proposed the MedNeRF model for synthesizing CT projections, but it can only generate new-view CT projections without reconstructing the internal details of the CT. Different from above methods, We build a 3D diffusion model for explicit CT reconstruction via denoising sampling, including both external contours and internal details.

% Our approach uses a diffusion model to learn the distribution of explicit CT volumes and uses viewpoint guidance to achieve a balance between fidelity and perceptibility.

\begin{figure*}[h]
    \centering
    \includegraphics[width=\textwidth]{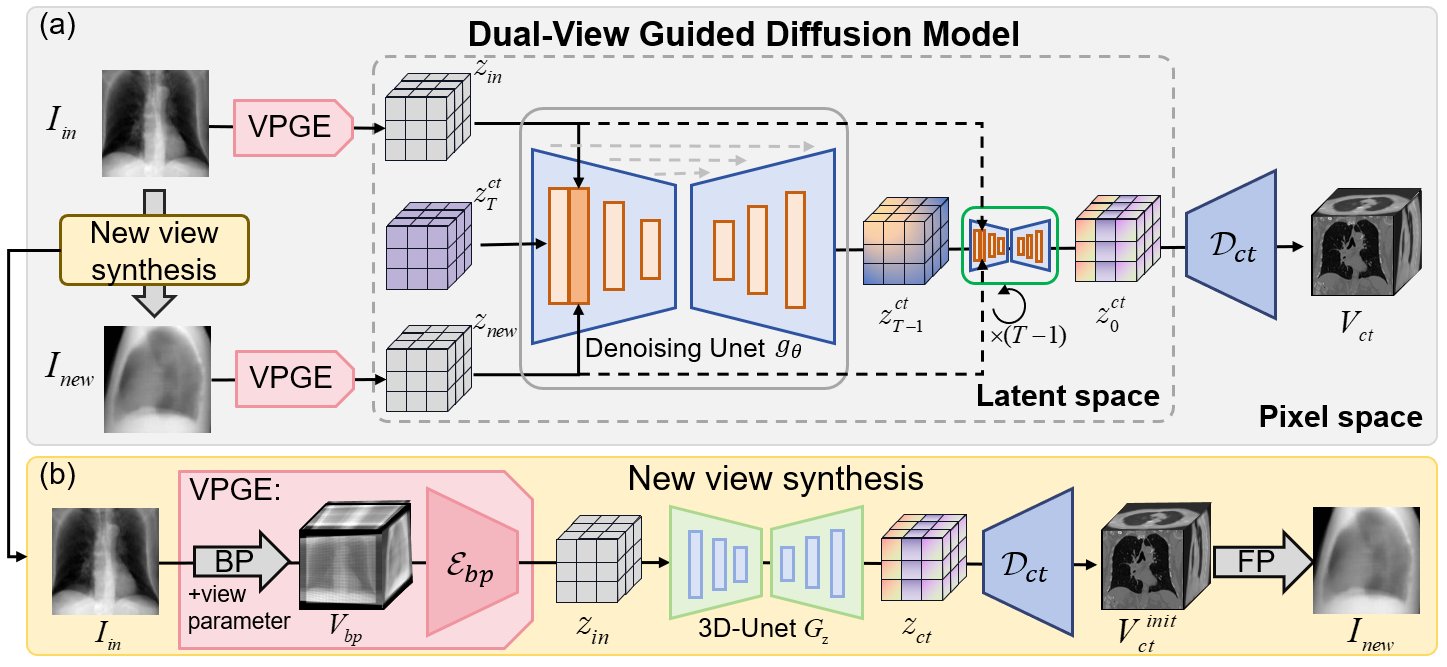}
    \caption{Framework of the proposed DVG-Diffusion.
    (a) We utilize dual-view (i.e., the real input view $I_{in}$ and the synthesized new view $I_{new}$) to jointly guide the reconstruction of the finer CT volume ($V_{ct}$). $I_{in}$ and $I_{new}$ are used as conditions to guide diffusion models to restore more detail.
    (b) We synthesize the new view ($I_{new}$) by predicting the initial CT volume from the input view ($I_{in}\to V_{ct}^{init}$) and then forward projecting the predicted CT to the new view X-ray ($V_{ct}^{init}\to I_{new}$). A core design is our view parameter guided encoder (VPGE). To more easily align the feature space and learn the mapping from 2D images to 3D CT volume, in our VPGE, the input X-rays are first back projected into 3D space and then encoded into latent space ($I_{in} \to V_{bp} \to z_{in}$). 
    }
    %Framework of the proposed DVG-Diffusion. First, $I_{in}$ is back-projected ($BP$) by initial view guidance and encoded independently with the same VQGAN as CT, which construct a pair of aligned latent spaces. In novel view synthesis stage, we learning the initialbase CT volume and generating novel view by forward projection ($FP$). In view guidance reconstruction stage, we simultaneously utilize both the real and novel view to guide diffusion to generate a fine CT volume. All learning processes are performed in the constructed aligned latent space.}
    \label{fig:x2ct framework}
\end{figure*}

\section{Proposed Method}
Compared to a single X-ray image, biplanar X-rays can reconstruct higher quality CT volumes. This superiority stems from introducing additional views. These views provide complementary information and facilitate the transition from 2D X-rays to 3D CT volumes. Inspired by this, we propose a dual-view guided diffusion model (DVG-Diffusion) for CT reconstruction from X-rays (see Fig. \ref{fig:x2ct framework}). The ``dual view guided'' in our paper has two meanings. The first is reflected in view parameter guidance in feature encoder and view guided diffusion model in CT reconstruction. The second is manifested in using real-view and generated new-view to guide CT reconstruction. This dual view guided mechanism can achieve a good balance between high fidelity and high perceptual quality for CT reconstruction.

\subsection{Overall Framework}
Our DVG-Diffusion consists of three steps (see Fig. \ref{fig:x2ct framework}(a)). (1) Feature extraction: 
%To more easily align features in the mapping from 2D images to 3D CT volume, 
We propose a view parameter-guided encoder (VPGE) to extract latent features $z_{in}$ and $z_{new}$ from the input view $I_{in}$ and new view $I_{new}$. (2) Feature transformation: We concatenate the features $z_{in}$ and $z_{new}$ as conditions for the diffusion model \cite{ho2020denoising}, using these dual view features to collaboratively guide the inverse generation of the latent representation of CT (donated as $z_0^{ct}$).
(3) Feature decoding: We decode $z_0^{ct}$ into pixel space to reconstruct the CT volume (donated as $V_{ct}$). For new view synthesis in Fig. \ref{fig:x2ct framework}(b), we synthesize the new view $I_{new}$ by predicting the initial CT volume $V_{ct}^{init}$ from the input real view $I_{in}$ and then forward projecting $V_{ct}^{init}$ to the new X-ray view $I_{new}$ (the details in Section~\ref{section2_4}). 
%  Pulling in the distance between the real image and the target domain makes the generated image closer to the source domain image.

\subsection{Feature Extraction}

Directly mapping 2D X-ray images to 3D CT volumes is ill-posed due to the lack of one-to-one correspondence between modalities. Unlike previous Diffusion-based  methods~\cite{bhunia2023person,tseng2023consistent,yang2023paint} that directly encode input images end-to-end, we propose a novel \textbf{view parameter guided encoder} (VPGE) to better exploit the inherent structure and geometrical relationships between two modalities. In VPGE, the input X-rays are first back-projected into 3D space and then encoded into latent space to more easily align and learn the mapping from X-ray to CT. This process involves the following two steps:

\begin{figure}[h]
    \centering
    \includegraphics[width=\linewidth]{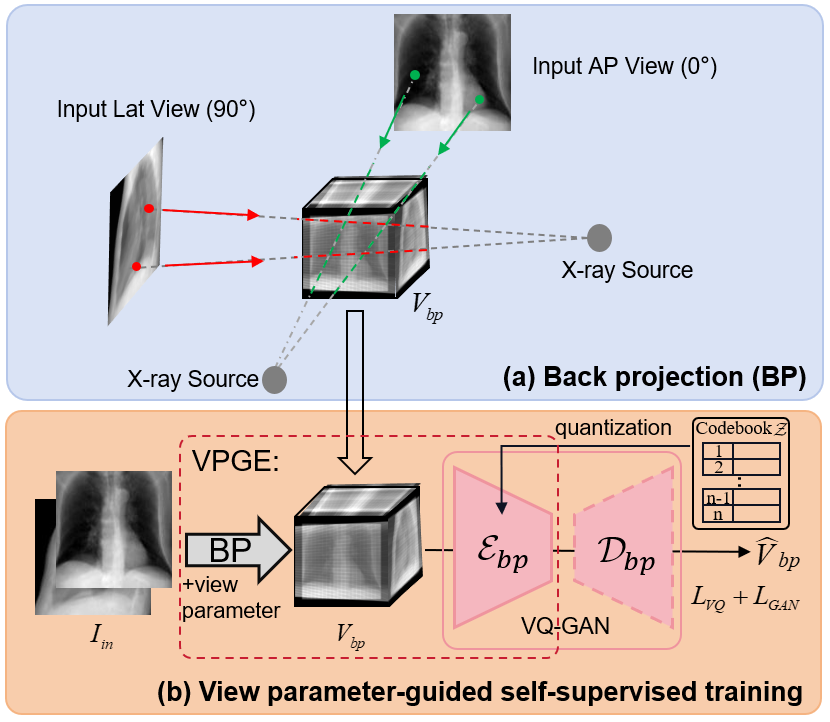}
    \caption{The detailed process of back projection and VPGE using biplanar inputs as an example. The backprojector (BP) unfolds the points on each 2D projection into 3D space separately and superimposes them to obtain a geometrically consistent 3D image $V_{bp}$. The robust discrete feature encoder $\mathcal{E}_{bp}$ is then obtained by self-supervised training of VQ-GAN.
    }
    \label{fig:bp}
\end{figure}

1) We back-project a given 2D X-ray image $I_{in}\in \mathbb{R}^{H\times W} $ into 3D space ($\mathbb{R}^{H\times W\times D}$) to obtain $V_{bp}$. $V_{bp}=BP(I_{in},M)$, where $BP$ is the backprojector from \cite{peng2021xraysyn}. It can properly place the X-ray image in 3D space while maintaining geometric consistency:
\begin{equation}
\label{eq-V-bp}
BP(I_{in},M):= \left\{
	\begin{aligned}
	&I_{in}/(\lvert L\rvert \Delta p),& \quad &My\in L,&\\
	&0,& \quad &otherwise,&\\
	\end{aligned}
	\right.
\end{equation}
where $M$ represents the view parameter matrix controlling rotation and translation, $L$ denotes line geometry constraints, $\Delta p$ denotes sample step, and $y$ denotes the coordinate point in 3D CT volume. 
%Following \cite{shen2022geometry}, 
To extend the backprojector to accept multi-view inputs, we back-project each input image separately and sum them together:
\begin{equation}
V_{bp}=\sum_{i=0}^n V_{bp}^i =\sum_{i=0}^i BP(I_{in}^i,M_i),
\end{equation}
where $n$ is the total number of input views, and $I_{in}^i$ denotes the $i$-th input image. An example of a visual back-projection for biplanar inputs is presented in Fig. \ref{fig:bp}(a).
%This backprojector process incorporates the physical view information of the input X-ray image into the encoding process for better alignment of the representation space.
This backprojection process incorporates the physical view information of the input X-ray images into the encoding process, which allows for better alignment of the representation space.

% The above details the encoding/decoding process for X-ray images. For CT encoding/decoding, we use a similar VQGAN approach without the $BP$ projection step. 

% utilizing a encoder specialized for 3D CT volumes via a self-supervised VQGAN

2) %By exploiting the powerful feature representation capabilities of self supervised learning, 
To fully exploit the inherent structure and geometrical relationships between X-ray and CT modalities, we employ a specialized encoder for the back-projected X-ray (donated as $V_{bp}$). This encoder, based on a self-supervised VQGAN~\cite{esser2021taming}, transforms $V_{bp}$ into a latent space, represented as $z_{in}$, where $z_{in}=\mathcal{E}_{bp}(V_{bp})$. We uniformly use $\mathcal{E}_{bp}(\cdot)$ to denote the CNN encoder $E(\cdot)$ and quantization $q(\cdot)$.
%To fully exploit the inherent structure and geometrical relationships between the X-ray and CT modalities, we use an encoder specialized for the back-projected X-ray ($V_{bp}$) via a self-supervised VQGAN~\cite{esser2021taming}, to transform $V_{bp}$ into latent space ($z_{in}=\mathcal{E}_{bp}(V_{bp})$).

The detailed flow of the encoder is shown in Fig. \ref{fig:bp}. Specifically, we encode the $V_{bp}$ 
as a latent code $\hat z_{in}$ using $\hat z_{in} = E(V_{bp}) \in \mathbb{R}^{(H/f)\times (W/f)\times (D/f)\times n} $, where $n$ is the dimension of the codes, and $f$ denotes a compression factor. Next, we perform element-wise quantization $q(\hat z_{in})$ on each code in $\hat z_{in}$, mapping it to a pre-learned discrete codebook $\mathcal{Z}=\left\{z_{k}\right\}_{k=1}^K$ as:
\begin{equation}
z_{in}=q(\hat z_{in})=(\mathop{\text{arg min}}_{z_{k}\in\mathcal{Z}}\Vert \hat z_{in}^{ij}-z_{k} \Vert)\in \mathbb{R}^{h*w*d*n},
\end{equation}
where $h=H/f, w=W/f, d=D/f$. $\hat z_{in}^{ij}$ denotes pixel value of row $i$ and column $j$ on $\hat z_{in}$. 
The reconstruction of $\hat V_{bp} \approx V_{bp}$ is performed by a CNN decoder $\mathcal{D}_{bp}$ based on the $z_{in}$, i.e., $\hat V_{bp}=\mathcal{D}_{bp}(z_{in})=\mathcal{D}_{bp}(q(\hat z_{in}))$.

In summary, our feature extraction stage employs the proposed VPGE to extract robust latent space features $z_{in}$ and $z_{new}$ from the input view $I_{in}$ and new view $I_{new}$, respectively.

\subsection{Feature Transformation and Decoding}

To efficiently process CT volume features and enhance their perceptual quality, we train a denoising UNet $g_\theta$ to learn the target latent representation of CT (donated as $z^{ct}_{0}$). 
%learn the latent representation of CT ($z^{ct}_{0}$) by training a denoising UNet ($g_\theta$). 
The network $g_\theta$ uses the features of input and new views ($z_{in}$ and $z_{new}$) as conditions to stepwise guide and transform $z^{ct}_{T}$ (pure Gaussian noise) into $z^{ct}_{0}$ (clear CT representation):
\begin{equation}
z^{ct}_{t-1}=\frac{1}{\sqrt{\alpha_{t}}}(z^{ct}_{t}-\frac{1-\alpha_{t}}{1-\overline{\alpha}_{t}}(z^{ct}_{t}-\sqrt{\overline{\alpha}_{t}}g_{\theta}(z^{ct}_{t},z_{in},z_{new},t))). \label{Eq:1}
\end{equation}
Where $t$ is time step and uniformly sampled from $\left\{1,2,\dots,T\right\}$, $z^{ct}_{t}$ can be obtained by adding Gaussian noise to $z^{ct}_{0}$ in the forward process with a fixed cosine schedule \cite{ho2020denoising}:
\begin{equation}
q(z^{ct}_{t}|z^{ct}_{t-1})\sim\mathcal{N}(z^{ct}_{t};\sqrt{\alpha_{t}}z^{ct}_{t-1},(1-\alpha_{t})\textbf{I}).
\end{equation}
The denoising network $g_\theta$ is trained for estimating $z_0^{ct}$, and supervised by Mean Square Error (MSE) loss:
\begin{equation}
L_{diff}:= \Vert g_{\theta}(z^{ct}_{t},z_{in},z_{new},t)-z_0^{ct}\Vert^2_{2}.
\end{equation}
With iterative sampling in Eq.~\ref{Eq:1}, we obtain the CT latent representation $z^{ct}_0$. Subsequently, the CT volume $V_{ct}$ is reconstructed by decoding $z^{ct}_0$ into pixel space by a decoder $\mathcal{D}_{ct}$ specialized for 3D CT volumes via VQGAN ~\cite{khader2022medical}.

%by the well-trained VQGAN-decoder.
Using diffusion models for CT reconstruction can substantially enhance perceptual performance. However, the stochastic nature of the denoising process may result in the generated samples lacking fidelity. 
In our method, the feature $z_{new}$ is extracted from the new view $I_{new}$, which is mined completely from the input X-ray image to improve data utilization efficiency. 
This provides stronger spatial constraints for the denoising process, facilitating the reconstruction of finer CT volumes with 3D consistency.
%fully exploit the potential 3D features of the input X-ray image for CT reconstruction with 3D consistency.
In the following section, we describe the process of synthesizing new view $I_{new}$ from input view $I_{in}$.

\subsection{New X-ray View Synthesis}\label{section2_4}

To obtain X-ray images with more 3D consistency, our new view synthesis is not achieved through a simple 2D-to-2D mapping. Instead, it relies on the VPGE module to leverage the 3D information from CT volume, enabling a more reliable 2D-to-3D-to-2D mapping.

Given an input X-ray $I_{in}$, our VPGE-based new view synthesis utilizes the following image processing pipeline:
\begin{equation}
I_{in}\xrightarrow {BP}V_{bp} \xrightarrow {\mathcal{E}_{bp}(V_{bp})} z_{in} \xrightarrow {G_z(z_{in})} z_{ct} \xrightarrow {\mathcal{D}_{ct}(z_{ct})} {V_{ct}} \xrightarrow {FP} {I_{new}}. 
\end{equation}
$G_z$ is realized as a 3DUNet \cite{khader2022medical} to map the feature of input real views $z_{in}$ to the latent representation of initial CT volume (donated as $z_{ct}$), and is supervised by following loss:
\begin{equation}
L_{rec}:=\Vert G_{z}(z_{in})-z_{ct}^{gt}\Vert_{2}^{2}.
\end{equation}
$\mathcal{D}_{ct}$ is a decoder specialized for 3D CT volumes via VQGAN~\cite{khader2022medical} without $BP$ projection.
We employed the same VQGAN architecture for both CT encoding/decoding and $V_{bp}$ encoding/decoding, ensuring spatially coherent representation between two modalities.
%a similar VQGAN without the BP projection.

We obtain the new view by $I_{new}=FP(V_{ct}, M_{new})$, where $M_{new}$ represents the parameter matrix of the new view $I_{new}$,
and $FP$ is a differentiable forward projector \cite{peng2021xraysyn} for CT-to-X-ray: 
{
\small
\begin{equation}
\label{Eq_FP}
FP(V_{ct},M_{new}):= \int V_{ct}(M_{new}^{-1}p)dL\approx \textstyle\sum_{p\in L} V_{ct}(M_{new}^{-1}p)\Delta p.
\end{equation}
}

\begin{figure*}[h]
    \centering
    \includegraphics[width=\textwidth]{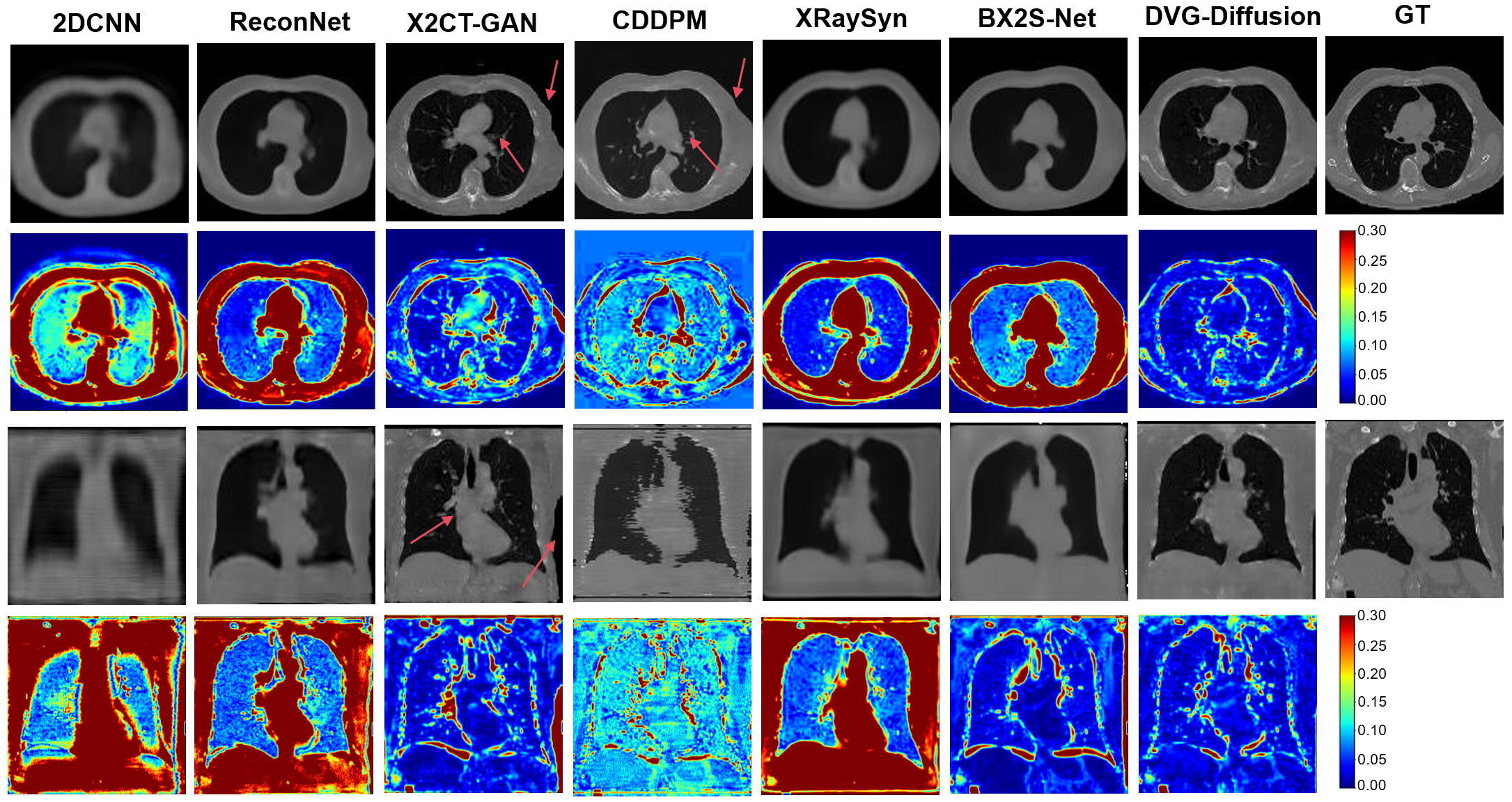}
    \caption{Example of reconstructed CT slices from biplanar X-rays using the proposed DVG-Diffusion and competing methods, along with the prediction error maps. Two sets of samples are slices along the axial plane (AXI) and coronal plane (COR). Red arrows in the figure indicate instances of contour distortion.}
    \label{fig:quantitative results0.3 }
\end{figure*}

Our view synthesis network captures features from X-rays aligned as closely as possible with CT via VPGE, facilitating the reconstruction of initial CT volumes with better 3D consistency. It synthesizes a reliable new view $I_{new}$ that is then fed into dual view guided diffusion model for finer CT reconstruction.

\section{Experiments and Results}
In this section, we first introduce the experimental implementation details. Then, we compare our DVG-Diffusion with several state-of-the-art methods and conduct ablation studies to validate our model.

\subsection{Experimental Setup}

\textbf{Dataset:} We evaluated our method on the LIDC-IDRI dataset \cite{armato2011lung}, containing 1018 chest CT volumes. Following~\cite{peng2021xraysyn}, we synthesized corresponding X-ray images of different views from real CT volumes using the differentiable forward projector (FP)~\cite{peng2021xraysyn}. We randomly selected 916 CT volumes for training, and used the remaining 102 volumes for testing. We followed the pipeline introduced in \cite{ying2019x2ct} for CT volume pre-processing.

\textbf{Implementation Details:} Two sets of experiments were conducted: (1) CT reconstruction from single-view X-ray, where we used Anterior-Posterior (AP) view, and (2) CT reconstruction from biplanar X-rays, where we used both Anterior-Posterior (AP) and Lateral (Lat) views.
We implemented our models using PyTorch on a single NVIDIA 4090 24GB GPU, with a batch size of 2 and the Adam optimizer \cite{kingma2014adam}. 

The implementation of different stages in our model is as follows: \textbf{(1) For VQGAN:} For CT encoding/decoding, the learning rate was set to $3\times10^{-4}$, and training was conducted over 100 epochs. In contrast, for $V_{bp}$ encoding/decoding, the learning rate was the same, but the training epoch was set to 50. \textbf{(2) For diffusion model:} We adopted the 3D UNet from \cite{khader2022medical} as our denoising model. We set the total diffusion timesteps to 100 and used a learning rate of $2\times10^{-4}$. \textbf{(3) For new view synthesis:} For CT reconstruction from single view X-ray, we used the input AP view (set to 0°) to synthesize Lat View (90°). For biplanar X-rays, we used the input AP view and Lat view to synthesize the middle (45°) view.

\textbf{Comparison Methods and Evaluation Metrics:} We compared our DVG-Diffusion with four CNN-based approaches (2DCNN \cite{henzler2018single}, ReconNet \cite{shen2019patient}, XRaySyn \cite{peng2021xraysyn} and BX2S-Net \cite{chen2023bx2s}), one GAN-based approach (X2CT-GAN \cite{ying2019x2ct}), and one conditional diffusion model (CDDPM \cite{saharia2022palette}). In particular, for the XRaySyn \cite{peng2021xraysyn} model, we evaluated the CT reconstruction network (3DPriorNet). CDDPM \cite{saharia2022palette} is a conditional diffusion model with our own implementation, conditioned on the back-projection of the X-ray image.
%In parallel, we included the model that generates $V_{ct}^{init}$ in the new view synthesis stage (called DVG-Net) as our another model setup for comparison.

We evaluated the performance of these methods with two distortion metrics (Peak Signal-Noise Ratio (PSNR), Structured Similarity Index (SSIM)), and one perceptual metric (Learned Perceptual Image Patch Similarity (LPIPS) \cite{zhang2018unreasonable}). PSNR and SSIM primarily assess overall image quality, whereas LPIPS evaluates perceptual details at the feature level.

\subsection{Comparison Results}
\begin{figure*}[h]
\centering
    \includegraphics[width=0.9\textwidth]{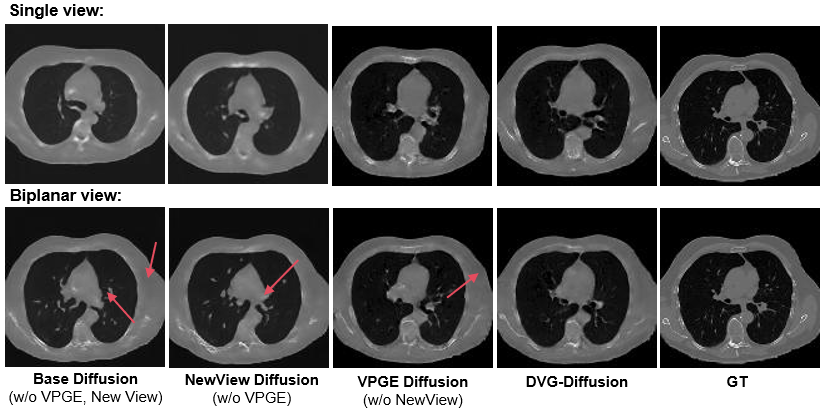}
    \caption{Visual results of baseline and the proposed DVG-Diffusion in ablation studies, using single-view and biplanar X-rays as inputs.}
    \label{fig:ablation}
\end{figure*}
We present the comparison results in Fig.~\ref{fig:quantitative results0.3 } and Table~\ref{table:quantitative}. 
It is widely recognized that CNN-based methods typically excel in generating images with good distortion metrics, while GAN-based methods are adept at capturing perceptual details \cite{ying2019x2ct}. Achieving a balance between perceptual and distortion metrics in X-rays-to-CT generation remains a challenging task. 
As shown in Table~\ref{table:quantitative} and Fig.~\ref{fig:quantitative results0.3 }, among all the compared baselines, the CNN methods XRaySyn and BX2S-Net deliver the top results on the distortion metric SSIM, despite their perceptual quality being subpar. The GAN-based X2CT-GAN performs well in the perceptual metric LPIPS but struggles with maintaining distortion accuracy.
This trade-off is also evident in Fig.~\ref{fig:quantitative results0.3 } (marked with red arrows), where the contour structure of the CT image generated by X2CT-GAN exhibits significant distortion. Furthermore, samples produced by the CNN method appear too smooth and lack detail, resulting in a substantial gap in overall quality compared to the ground truth, as shown by the error maps. 
In contrast, our proposed DVG-Diffusion method successfully achieves a good balance between high fidelity and perceptual quality in CT reconstruction, resulting in the best performance among all competing methods in both perceptual and distortion metrics.

\begin{table*}[t]
  \centering
  \caption{Quantitative comparison of five CT reconstruction methods using Single-view X-ray and Biplanar X-rays as inputs, respectively. \textbf{Bolding} and \underline{underline} denote the best and the second best result, respectively.}
  \label{table:quantitative}
  %    \normalsize
  \begin{tabular}{cccc|ccc}
  \hline
  \multirow{2}{*}{\textbf{Method}} & \multicolumn{3}{c}{\textbf{Single-view X-ray}}  & \multicolumn{3}{|c}{\textbf{Biplanar X-rays}}\\
    \cline{2-7}
    % \multirow{1}{*}{Metrics}
    & \textbf{PSNR $\uparrow$} & \textbf{SSIM $\uparrow$} & \textbf{LPIPS $\downarrow$}   & \textbf{PSNR $\uparrow$} &\textbf{SSIM $\uparrow$} &\textbf{LPIPS $\downarrow$}   \\
    \hline
     2DCNN \cite{henzler2018single}  &  19.01&  0.480&  49.71& 22.85& 0.595&46.05\\
    % \hline
     ReconNet \cite{shen2019patient}& 21.35 & 0.521& 46.53& 24.01& 0.625 &40.70\\
%     X2CT-CNN \cite{ying2019x2ct} &  22.18&  \textbf{0.553} &  44.27 & 26.34& \textbf{0.682}&37.29\\
     X2CT-GAN \cite{ying2019x2ct} &  21.40 &  0.505 & 39.62 & 24.82 & 0.613&31.73\\
 XRaySyn \cite{peng2021xraysyn} & 21.07& \textbf{0.558} & 46.20 & 24.45& \textbf{0.684}&39.43\\

 CDDPM \cite{saharia2022palette} &20.05 &0.508 &46.95 & 22.04& 0.610&41.16\\
 BX2S-Net \cite{chen2023bx2s} & 21.19& \textbf{0.558} & 44.65& 24.69& 0.658&38.88\\
 \hline
  % DVG-Net (Ours) & \textbf{23.08}& \textbf{0.568}& \underline{38.59} & \textbf{27.25}& \textbf{0.702}& \underline{29.50}\\
  DVG-Diffsion (Ours) & \textbf{22.54} & \underline{0.541} & \textbf{36.79} & \textbf{26.84} & \underline{0.679} & \textbf{27.35}\\
    \hline
  \end{tabular}
\end{table*}

\begin{table*}[t]
  \centering
  \caption{Ablation study to evaluate the impact of VPGE and new view synthesis guidance modules in DVG-Diffusion, using single view and biplanar X-rays as inputs.}%  while SP denotes single plane.}
  \label{table:ablation}
% \resizebox{\linewidth}{8mm}{
  \begin{tabular}{c|c|c|c|c|c|c|c}
    \hline
   
       & VPGE & New View 
 & Single&Biplanar& PSNR ↑  &SSIM ↑ & LPIPS ↓\\
       \hline

   Base-Diffusion&-& - & \checkmark  && 20.05& 0.508& 46.95\\
NewView-Diffusion&-& \checkmark  & \checkmark  &&20.62&0.519& 44.67\\

 VPGE-Diffusion&\checkmark & -
 & \checkmark  && 22.29& 0.532 &37.59\\
DVG-Diffusion (Ours) &\checkmark &\checkmark  & \checkmark  &&  \textbf{22.54}&\textbf{0.541}&\textbf{36.79}\\ %\hline
 \hline
 Base-Diffusion& -& - & & \checkmark  & 22.04& 0.610&41.16\\
 NewView-Diffusion& -& \checkmark  & & \checkmark  & 22.53& 0.625&39.85\\
 VPGE-Diffusion& \checkmark & -
& & \checkmark  & 26.47 & 0.667 &28.11\\
  DVG-Diffusion (Ours) & \checkmark & \checkmark  & & \checkmark  & \textbf{26.84}& \textbf{0.679}&\textbf{27.35}\\
        \hline
  \end{tabular}
  % }
\end{table*}

\subsection{Results of Ablation Experiments}\label{section:ablation}

This section investigates the effects of VPGE and the new view synthesis guidance in DVG-Diffusion. We remove each of these modules from the DVG-Diffusion model individually and evaluate the resulting variants: Base-Diffusion (both VPGE and new view synthesis guidance removed), NewView-Diffusion (VPGE removed), and VPGE-Diffusion (new view synthesis guidance removed). All derived models were trained using a strategy similar to our DVG-Diffusion. Results are presented in Table~\ref{table:ablation} and Fig.~\ref{fig:ablation}.

\textbf{Effectiveness of VPGE:} Directly mapping 2D X-ray images to 3D CT volumes is inherently ill-posed due to the lack of one-to-one correspondence between the two modalities. This challenge is evident in the lowest performance of the Base-Diffusion variant (see Table~\ref{table:ablation}). By incorporating the VPGE module into the Base-Diffusion model, we simplify the decomposition of the original complex end-to-end 2D to 3D mapping. VPGE pre-codes the X-ray images and constructs a more feasible 3D to 3D latent spatial mapping, which leads to a significant improvement in performance (comparing Base- Diffusion and VPGE-Diffusion).

\textbf{Effectiveness of new view synthesis guidance:} Removing the new view synthesis guidance from our DVG-Diffusion method results in a noticeable decrease in image fidelity, as evidenced by the SSIM distortion metric, which drops from 0.679 in our DVG-Diffusion with biplanar inputs to 0.667 in the VPGE-Diffusion variant. This reduction is also evident in Fig.~\ref{fig:ablation} (marked by red arrows), where the image generated by the VPGE-Diffusion variant exhibits noticeable distortion in the contour structure. In contrast, the use of our proposed new view synthesis guidance greatly improves image fidelity by providing supplementary new view X-ray images as guidance.

\section{Discussion}
\subsection{Impact of X-ray Quantity and Angular Distribution on CT Reconstruction}\label{setionD:Few-view reconstruction}
In our experiments, we tested the performance of our DVG-Diffusion method in two experimental settings: CT reconstruction from single-view X-ray and biplanar X-rays. Intuitively, using more input views can lead to better 3D CT reconstruction. However, increasing the number of input X-ray views also introduces additional costs, such as increased acquisition time and computational complexity. Selecting the optimal parameters that balance reconstruction quality and cost is a crucial challenge. Moreover, for a given number of input views, the performance of CT reconstruction may vary depending on the angular distribution of the views.

In this section, we conduct extensive experiments to explore the impact of the number of input views and angular distribution on CT reconstruction quality. Specifically, we conducted a series of experiments with different numbers of input views, ranging from 1 to 20. We uniformly sampled the expected number of input X-ray views from two view ranges: 0°-90° and 0°-360°, with the first input view set as the Anterior-Posterior (AP). These ranges, 0°-90° and 0°-360°, are two common sampling ranges for CT reconstruction from X-ray, referred to as limited angle CT and sparse-view CT \cite{chung2023solving}.  To accurately evaluate the performance of view quantity and view selection on CT reconstruction, we removed the new-view synthesis module in DVG-Diffusion to eliminate the influence of synthesized new views on the final performance. Moreover, we compared our method with one of the SOTA CT reconstruction methods, ReconNet \cite{shen2019patient}, to validate the superiority of DVG-Diffusion. Other comparison methods listed in Table \ref{table:quantitative} were specifically designed for single-plane or biplanar X-rays and therefore are not involved in this comparison.

\begin{figure*}[h]
    \centering
    \includegraphics[width=\textwidth]{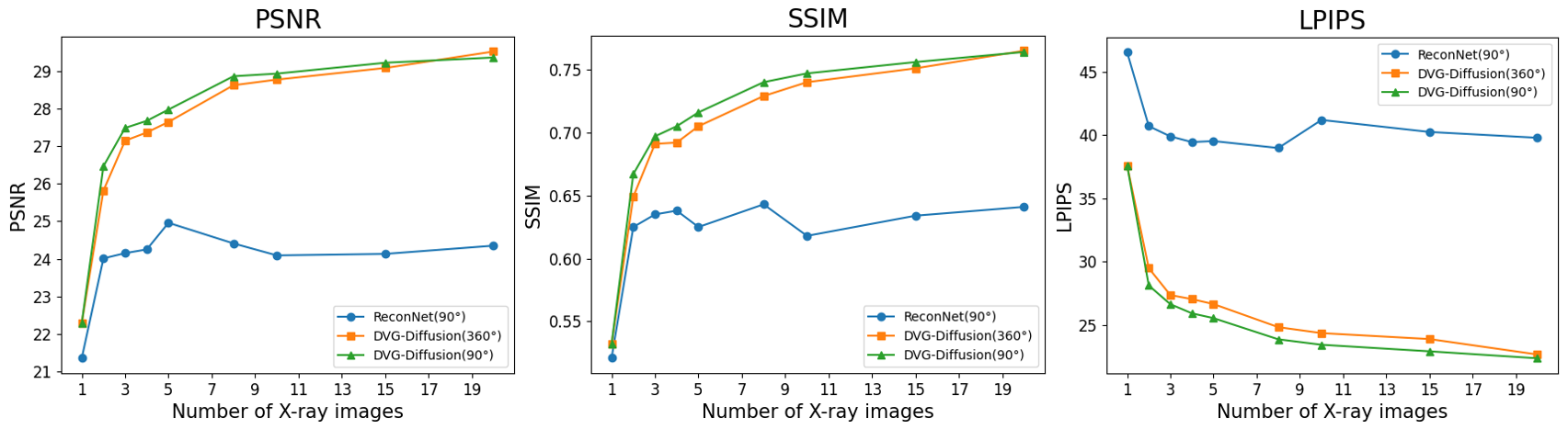}
    \caption{Performance curves of CT reconstruction using DVG-Diffusion (without new view synthesis) for different numbers of input views and different angle sampling ranges, and a comparison with ReconNet.}
    \label{fig:mutil_m}
\end{figure*}

\begin{figure*}[t]
    \centering
    \includegraphics[width=\textwidth]{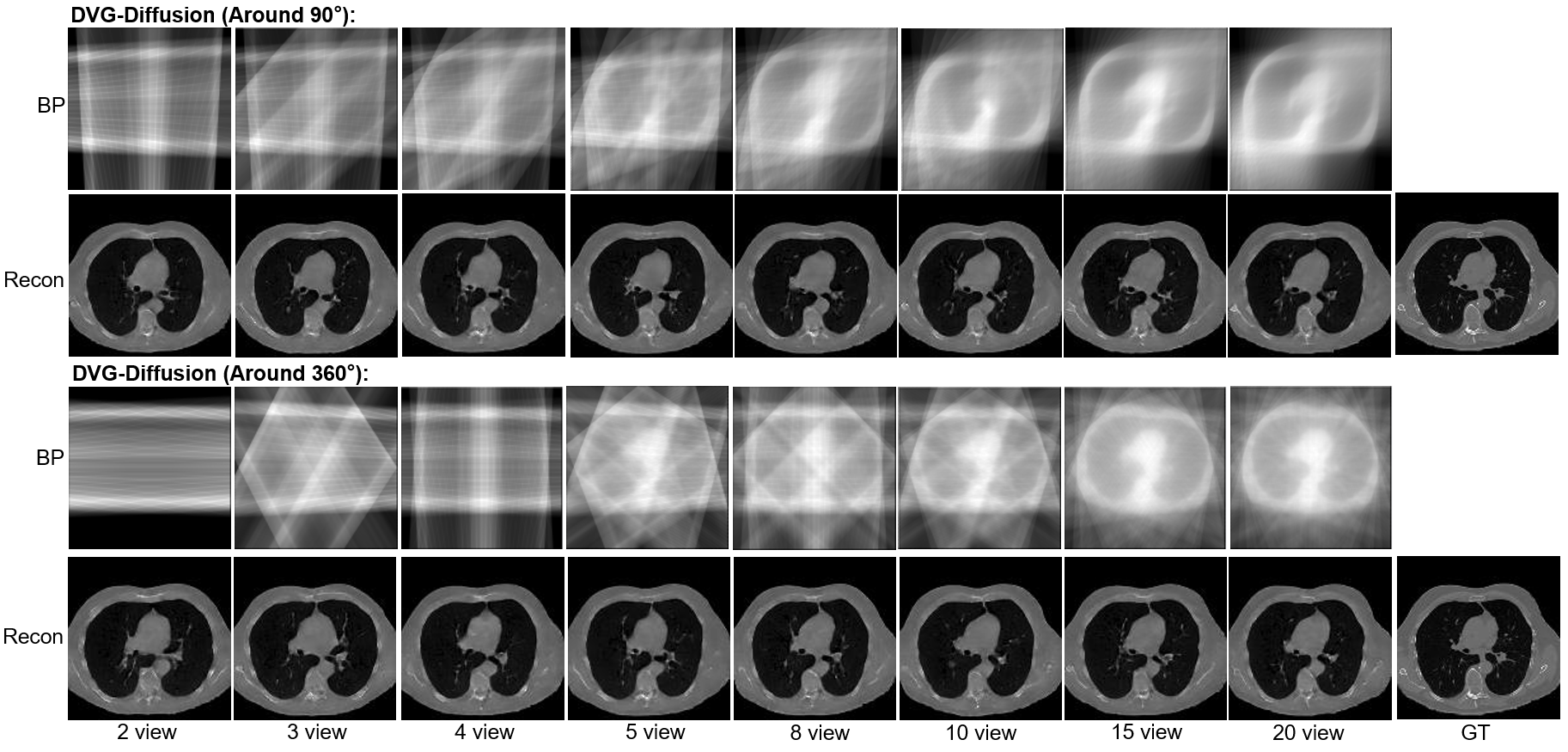}
    \caption{Visual results of CT reconstruction using DVG-Diffusion (without new view synthesis) for different numbers of input views and different angle sampling ranges.}
    \label{fig:mutil-view}
\end{figure*}

\textbf{Observation \uppercase\expandafter{\romannumeral1}: Superior CT reconstruction performance with 90° sampling than 360° for few-view inputs.}
% A side-by-side comparison of DVG-Diffusion (around 360°) and DVG-Diffusion (around 90°) shows that uniform sampling is better over 90° than 360°. The reconstruction performance of DVG-Diffusion (around 360°) does not increase steadily with increasing viewing angle, and even decreases when the number of X-ray images goes from 4-5. DVG-Diffusion (around 90°), on the other hand, guarantees a steady increase in reconstruction quality. This is due to the fact that the imaging of the X-ray projection image is obtained by integrating the sampling points of the projected projection line, so that the same values are obtained in the front-back symmetric region. For the sparse projection reconstruction task, there is a need to try to ensure that each projection contains as much unique amount of information as possible, so sampling within 90° will outperform sampling within 360°. This also guides the angle at which the new view X-ray image is synthesised in our method.
Fig. \ref{fig:mutil_m} compares the performance of DVG-Diffusion for reconstructions using around 360° and around 90° views, showing that with fewer than 10 views, sampling around the 90° range yields better reconstruction results than sampling around the 360° range.
The improved performance can be attributed to the way in which the input X-ray image is generated, a process known as forward projection (FP). This process involves integrating samples along the projection line, resulting in similar information for symmetrical regions. Consequently, it is easier to sample duplicate information in 360° when the number of sampled input views is limited. X-ray images sampled within the orthogonal view range (90°) contain more comprehensive information, leading to enhanced performance compared to those sampled around 360°. This observation is further supported by Figure \ref{fig:mutil-view}, where images sampled from the 90° range reveal the basic shape contours of CT in back projection (BP) more rapidly than those sampled from the 360° range (see the 4-view examples).
%This is because the X-ray projected images are imaged by integrating sampled points along the projected line, so that only similar information can be obtained in the anterior-posterior symmetric region. Therefore, when the input view is limited, it is easy to sample duplicate information in 360°.
%Images sampled within the orthogonal view range (around 90°) contain more sufficient information and thus outperform those sampling around 360°.
%This can also be demonstrated in Fig. \ref{fig:mutil-view} that images sampled around 90° show the shape contours of CT in back projection (BP) faster (see 4 view).
Our new-view synthesis experiment is also based on these findings, suggesting that the angle for synthesizing new X-ray views should ideally be selected within the 90° range. 
%This also guides our method in which the angle for synthesizing new view X-ray images can be selected in the 90° range.
However, as the number of views increases, the 360° sampling gradually outperforms the 90° sampling. This occurs because the information within the 90° range becomes fully exploited, necessitating a broader range of spatial information for achieving a more detailed reconstruction.
%As the number of views increases, sampling around 360° gradually outperforms sampling around 90°, as the information around the 90° range is sufficiently mined, and therefore a larger range of spatial information is required for more fine-grained reconstruction.

\textbf{Observation \uppercase\expandafter{\romannumeral2}: Inflection points in reconstruction performance occur at 3 and 8 inputs.}
% A longitudinal comparison was conducted to evaluate the performance of DVG-Diffusion (around 90°) with varying numbers of inputs. The results clearly demonstrate that increasing the number of input X-ray images from one to two leads to a significant improvement, and a larger improvement is observed when increasing the number of input images from two to three. However, it is important to note that the performance indicators show only a marginal improvement when more than three input images are used, indicating that three input images are sufficient for optimal performance. We prioritize views with more unique information when generating new views, as orthogonal views provide the most unique information and the greatest performance improvement.
In this experimental section, we further explore the impact of different numbers of input views on the CT reconstruction performance, focusing on a 90° sampling range.  Fig. \ref{fig:mutil_m} indicates that the reconstruction performance of DVG-Diffusion displays a noticeable inflection point when the number of input views changes. 
In particular, inflection points in the reconstruction performance are observed at 3 and 8 inputs. When the number of input views is less than 3, we observe a rapid improvement in the reconstruction performance as each additional view is included. The information from these initial views is crucial, as they provide the fundamental structure and orientation details necessary for the reconstruction.
However, as the number of input views increases from 3 to 8, the rate of improvement in reconstruction performance begins to slow down. This is because the additional views provide increasingly overlapping or redundant information, which contributes less to the overall quality of the reconstruction.
%For better reconstruction, using 8 images strikes a balance between performance and cost. 
This experiment offers insights into the number of synthesized views required, which will be further discussed in Section \ref{section4.E}.
%This experiment provides guidance on the number of synthesized views needed, which we will further discuss in Section \ref{section4.E}.

%We compare the performance of DVG-Diffusion (around 90°) for different numbers of inputs. As shown in Fig. \ref{fig:mutil_m}, there are two inflection points when the input X-ray images are 3 and 8. When the input views are 3 or fewer, increasing the number of input views significantly improves the reconstruction performance, with the greatest improvement occurring with 2 views. Therefore, we conclude that 2 to 3 X-ray images are sufficient for effective CT reconstruction.
%For better reconstruction, using 8 images strikes a balance between performance and cost.
%This experiment provides guidance on the number of synthesized views needed, which we will further discuss in Section \ref{section4.E}.

\textbf{Observation \uppercase\expandafter{\romannumeral3}: Our method demonstrates robustness as the number of X-ray image inputs increases.}
We compare the performance of DVG-Diffusion with ReconNet, focusing on a 90° sampling range. As shown in Fig. \ref{fig:mutil_m}, our method leads across the board in fidelity (PSNR, SSIM) and perceptual (LPIPS) metrics as the number of inputs increases. We find that the CT reconstruction performance of ReconNet plateaus and does not improve further when the number of input views exceeds five. In contrast, our DVG-Diffusion model exhibits a steady enhancement in performance with an increasing number of input views. This observation demonstrates the effectiveness of the VPGE module and the efficient diffusion model learning strategy in our DVG-diffusion model. These designs enable our model to effectively handle different numbers of input views and obtain robust reconstruction quality. %Our VPGE module reduces the difficulty of end-to-end feature extraction, is more resilient to more complex inputs, and is better compatible with more inputs.
%Meanwhile, the ReconNet model no longer improves the distortion metrics and perceptual metrics when the number of inputs is greater than five, and fluctuates significantly. In contrast, our DVG-Diffusion model does not have this limitation, and the model performance steadily improves with increasing input views. It is also more evident from Fig. \ref{fig:mutil_m} that the DVG-Diffusion model has more stable performance as the number of input views increases. 

\begin{table}[h]
  \centering
  \caption{Quantitative results of CT reconstruction guided by synthesizing a new view with different angles when single-view X-ray input. (The input view is an AP (0°) view.) w/o denotes no new view input.}
  \label{table:different angle}
% \resizebox{\linewidth}{8mm}{
  \begin{tabular}{cccccccl}
    \hline
   
         &w/o &10° & 30°& 45° & 60°  &90° &120°\\
        \hline
 PSNR $\uparrow$  &22.29& 22.46& 22.47& 22.43&   22.48&\textbf{22.54} &22.47\\

       SSIM $\uparrow$  &0.532&0.537&0.539& 0.540&   0.539&\textbf{0.541} &0.537\\
 LPIPS $\downarrow$  &37.59&37.11&36.84& 37.08&   37.15&\textbf{36.79} &37.23\\
 \hline

  \end{tabular}
  % }
\end{table}

\begin{table}[h]
  \centering
  \caption{Experimental settings for new view synthesis based on different input views.}
  \label{table:new view angle}
% \resizebox{\linewidth}{8mm}{
  \begin{tabular}{c|c|c|c}
    \hline
   
       input view &single-view& two-view& three-view\\
        \hline
 input angle& 0°& 0°, 90°& 0°, 45°, 90°
\\

       1 new view &90°&45°& 22.5°\\
 2 new view &45°, 90°&22.5°,  67.5°& 22.5°,  67.5°\\
 \hline

  \end{tabular}
  % }
\end{table}

\begin{figure*}[h]
    \centering
    \includegraphics[width=\textwidth]{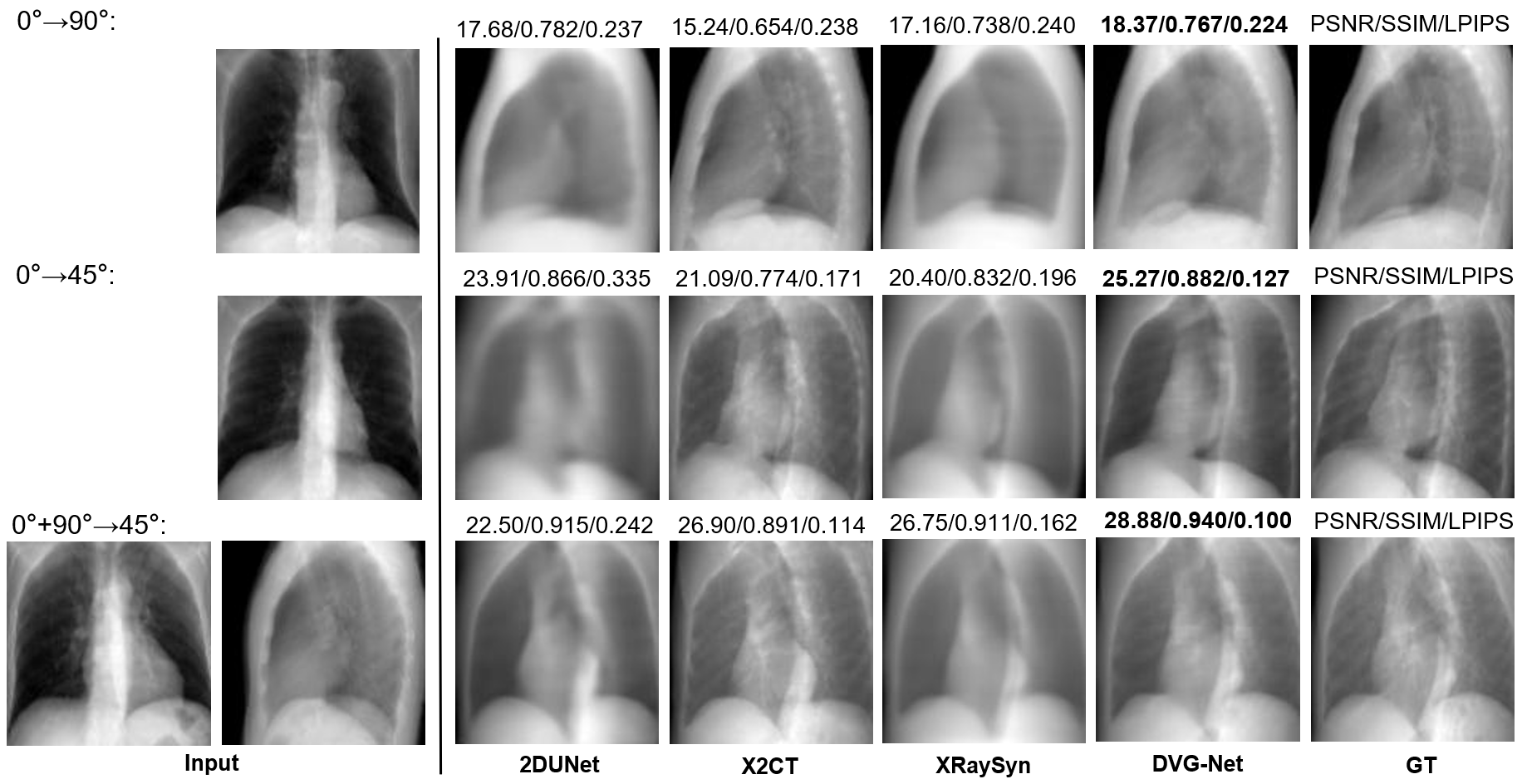}
    \caption{Visual results of new view synthesis using the proposed DVG-Diffusion (New View Synthesis Stage) and other comparison methods.}
    \label{fig:xray synthesis results}
\end{figure*}

\subsection{An In-Depth Analysis of the New View Synthesis}\label{section4.E}
% We evaluate the performance of the view generation network for new view synthesis. 
%In this section, we will evaluate the performance of the X-ray images generated by the new view synthesis stage in our model, referred to as DVG-Net. In addition, we evaluate the effect of CT reconstruction when synthesising different number or different angles of new views to further explore the effective scenes of new views.

% The superiority of our DVG-Diffusion model can be attributed to three key components: the efficient VPGE module, the innovative new view synthesis, and the robust dual-view guided diffusion model. Although we established the effectiveness of these individual components in Section \ref{section:ablation}, several questions regarding the new view synthesis remain to be answered. Specifically, when generating a new view, which angle would yield the most significant benefits for the final CT reconstruction? Furthermore, what is the optimal number of new views to be synthesized for the best CT reconstruction performance? 
% For example, given biplanar X-rays as input, should we synthesize 1, 2, or even more new views? In this section, we conduct extensive experiments to unravel the impact of the new view synthesis on CT reconstruction performance.
The superiority of our DVG-Diffusion model can be attributed to three key components: the efficient VPGE module, the innovative new view synthesis, and the robust dual-view guided diffusion model. We established the effectiveness of these individual components in Section \ref{section:ablation}. In this section, we conduct extensive experiments to unravel the impact of the new view synthesis on CT reconstruction performance. 
Specifically, we verified three key aspects: (1) the optimal angle for new view synthesis that yields the most favorable results for final CT reconstruction; (2) the ideal number of synthesized new views required to achieve the best CT reconstruction; and (3) the superior performance of new view synthesis in comparison to other methods.
%For example, given biplanar X-rays as input, should we synthesize 1, 2, or even more new views?

\textbf{Observation \uppercase\expandafter{\romannumeral4}: Synthesizing orthogonal view brings the best performance gains for a single view input.}
In our previous observation, we demonstrated superior CT reconstruction performance with 90° sampling than 360° for few-view inputs. This suggests that the angle for synthesizing new X-ray views should ideally be selected within the 90° range. Here, we further investigate the new view synthesis for a single view input. As shown in Table \ref{table:different angle}, the best CT reconstruction performance is achieved when generating the 90° views. Constructing orthogonal views (0° and 90°) leads to the greatest performance improvement, consistent with the study in Section \ref{setionD:Few-view reconstruction}. Despite the best performance achieved when generating orthogonal view, our experiments also confirm that, for a given single view as input, generating new views at any angle still provides some improvement.

%We show quantitative results of CT reconstruction when synthesising new views with different angles using single view input in table \ref{table:different angle}. It can be seen that the best CT reconstruction performance is achieved when generating the 90° views. Constructing orthogonal views (0° and 90°) leads to the greatest performance improvement, consistent with the study of the number of views in Section \ref{setionD:Few-view reconstruction}. While smaller angle views result in more realistic samples, they contain less valid 3D information than 90°. It is worth noting that the new views generated at all angles result in varying degrees of enhancement.
\begin{table*}[h]
  \centering
  \caption{Performance of CT reconstruction using different numbers of new views to guide CT reconstruction when different numbers of input views are used.}
  \label{table:new-view reconstruction}
% \resizebox{\linewidth}{8mm}{
  \begin{tabular}{c|ccc|ccc|ccc}
    \hline
   
        \multirow{2}{*}{New views}&\multicolumn{3}{c|}{single-view}& \multicolumn{3}{c|}{two-view}& \multicolumn{3}{c}{three-view}\\
        \cline{2-10}
 & \textbf{PSNR $\uparrow$}& \textbf{SSIM $\uparrow$}& \textbf{LPIPS $\downarrow$}& \textbf{PSNR $\uparrow$}& \textbf{SSIM $\uparrow$}& \textbf{LPIPS $\downarrow$}& \textbf{PSNR $\uparrow$}& \textbf{SSIM $\uparrow$}& \textbf{LPIPS $\downarrow$}\\
 \hline
    0&22.29&0.532& 37.59& 26.47& 0.667& 28.11& \textbf{27.49}& \textbf{0.697}& 26.63\\
 1&22.54&0.541& 36.79&\textbf{26.84}& \textbf{0.679}& \textbf{27.35}& 27.48 & 0.696 & \textbf{26.58}\\
 2& \textbf{22.59}& \textbf{0.543}& \textbf{36.57}& 26.80& 0.676& 27.51& 27.39& 0.696&26.75\\
  \hline
  \end{tabular}
  % }
\end{table*}

\textbf{Observation \uppercase\expandafter{\romannumeral5}: Enhanced performance of the new view synthesis with limited input views.}  We investigate the performance of our DVG-Diffusion model when given a limited number of X-ray views as inputs, such as a single view, biplanar views, and three views. For each scenario, we synthesize 1, 2, and 3 new views and assess their impact on CT reconstruction. As per Observation I, the angles of these new views are uniformly sampled within a 0-90° range, as detailed in Table \ref{table:new view angle}. The results, presented in Table \ref{table:new-view reconstruction}, reveal that when only one or two real views are available as input, our new view synthesis improves CT reconstruction performance. However, when the number of real views increases to three or more, the addition of synthesized views does not further enhance the performance. This observation aligns with our previous experimental findings in Fig. \ref{fig:mutil_m}, where the reconstruction performance improves substantially when three or fewer input views are provided and then adding real views, but the improvements taper off after three views. This suggests that when three or more input views are given, the model already obtains sufficient information from the existing real views, and the synthesized views do not contribute additional significant information. Based on these findings, we optimize our final model to synthesize one new view when given single-plane or biplanar inputs. No new views are synthesized when more planes are available as input. Despite the diminishing returns of our new view synthesis in CT reconstruction, our new view strategy is still quite useful in clinical scenarios, where typically only one or two X-ray views (Anterior-Posterior view (AP) and Lateral view (Lat)) are available. In cases where more views are available, our method still delivers excellent multi-view reconstruction performance through the effective coding capabilities of VPGE, as highlighted in Fig. \ref{fig:mutil_m}.

\begin{table}
  \centering
  \caption{Quantitative results of new view synthesis using the proposed DVG-Diffusion (new view synthesis stage) and other comparison methods.}
  \label{table:x-ray synthesis}
% \resizebox{\linewidth}{8mm}{
  \begin{tabular}{c|c|c|c|c|c}
    \hline
   
        && UNet &X2CT & XRaySyn& DVG-Net\\
        \hline
 \multirow{3}{*}{$0^{\circ}  \rightarrow 90^{\circ}$}& PSNR $\uparrow$& 18.57&17.77& 17.15&\textbf{19.08}\\

    &SSIM $\uparrow$&0.766&0.739& 0.753& \textbf{0.778}\\
 &LPIPS $\downarrow$&0.234&0.188& 0.238&\textbf{0.186}\\
  \hline
 \multirow{3}{*}{$0^{\circ}  \rightarrow 45^{\circ}$}& PSNR $\uparrow$
& 22.18&21.63& 21.45&\textbf{22.58}\\

  &SSIM $\uparrow$
&\textbf{0.843}&0.799& 0.824& 0.842\\
 &LPIPS $\downarrow$&0.294&0.154&0.196&  \textbf{0.148}\\ \hline
 % \multirow{3}{*}{\makecell[c]{$0^{\circ}+90^{\circ} \\ \rightarrow 45^{\circ}$}}& PSNR $\uparrow$
 \multirow{3}{*}{\makecell[c]{$0^{\circ} + 90^{\circ}$ \\ $\rightarrow 45^{\circ}$}} & PSNR $\uparrow$ 
& 25.78&26.03& 25.67&
\textbf{27.43}\\
 & SSIM $\uparrow$
& 0.911&0.883& 0.910&\textbf{0.922}\\
 
& LPIPS $\downarrow$& 0.240& 0.118 & 0.161&\textbf{0.099}\\
\hline

  \end{tabular}
  % }
\end{table}

\begin{figure*}[h]
    \centering
    \includegraphics[width=\textwidth]{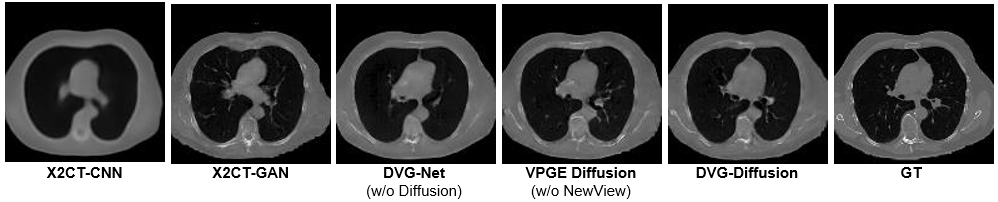}
    \caption{Visual results of CT reconstruction using CNN methods and generative methods (GAN, Diffusion).}
    \label{fig:dvg_net_diffusion}
\end{figure*}

\begin{figure*}[h]
    \centering
    \includegraphics[width=0.8\textwidth]{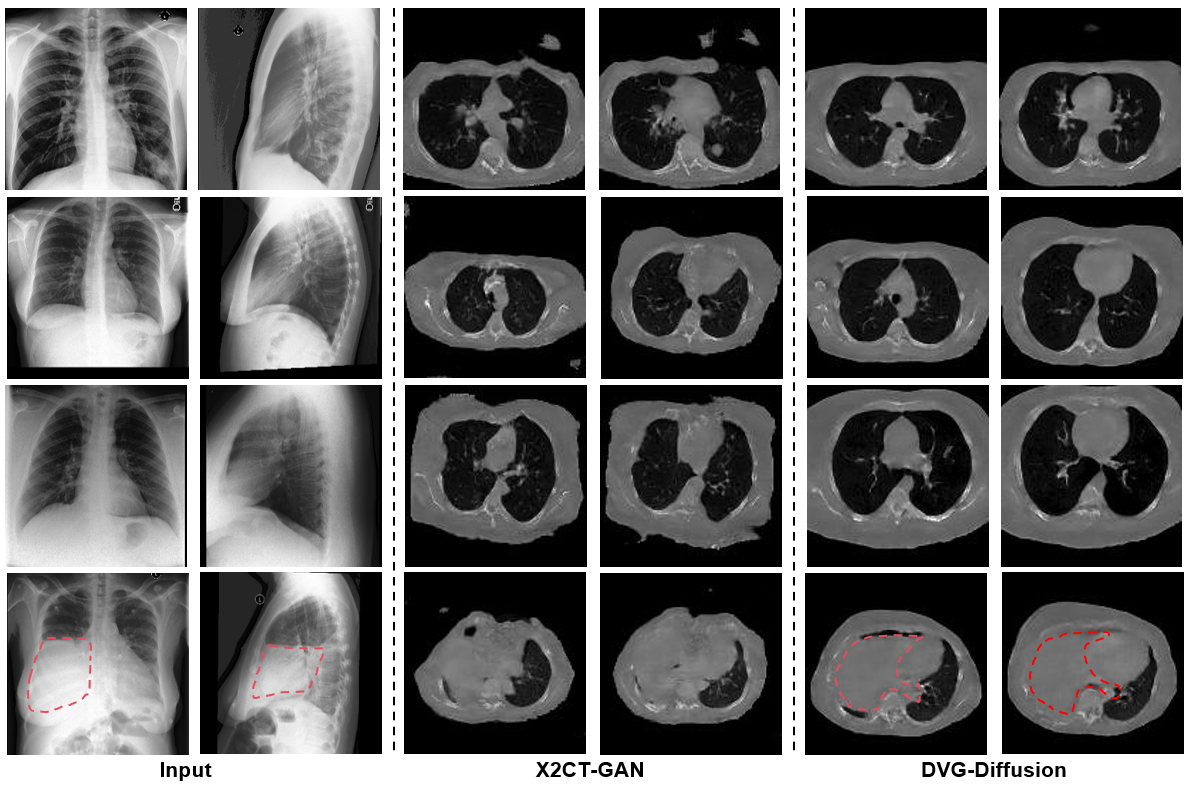}
    \caption{Visual results of CT reconstruction using real-world biplanar X-ray images.}
    \label{fig:real data}
\end{figure*}

\textbf{Observation \uppercase\expandafter{\romannumeral6}: Superior new view synthesis performance compared with other methods.}
Finally, we compare the performance of the X-ray images ($I_{new}$) generated by the new view synthesis stage in our DVG-Diffusion model (referred to as DVG-Net) with other state-of-the-art methods. In this comparison, we include one 2D method (UNet with residual blocks \cite{zhang2018road}), two 3D methods (X2CT \cite{ying2019x2ct} and XRaySyn \cite{peng2021xraysyn}). We present the results in Table \ref{table:x-ray synthesis} and Fig. \ref{fig:xray synthesis results}. The UNet method \cite{zhang2018road} is a 2D approach and learns the mapping between views directly, which limits its ability to capture and reconstruct reliable 3D information. As a result, the generated projections from UNet may lack essential depth and structural details. The X2CT method \cite{ying2019x2ct} aims to generate realistic samples using a GAN-based approach. While the overall appearance of the generated projections might be visually plausible, the method often struggles to maintain the fidelity of contours and intricate features, which are crucial for accurate CT reconstruction. The XRaySyn method \cite{peng2021xraysyn} tends to produce synthesized samples that appear overly smooth, causing them to miss essential details necessary for accurate reconstruction. This smoothing effect may lead to suboptimal CT reconstructions, as the algorithm struggles to model fine structures and subtle variations in the data. In contrast, our method consistently outperforms the other methods, as evidenced by the more accurate contours and detailed features in the synthesized projections. This superior performance highlights the effectiveness of our new view synthesis and its potential for improving CT reconstruction, especially in challenging scenarios with limited input views.

\subsection{Fidelity vs. Perceptual}
% As shown in Fig. \ref{fig:x2ct framework}, our framework can output two different CT volumes, including $V_{ct}^{init}$ (initial reconstructed) and $V_{ct}$ (final reconstructed). We compare DVG-Net and DVG-Diffusion in Table \ref{table:quantitative}. Compared to $V_{ct}$, which uses a generative diffusion model, $V_{ct}^{init}$ has a higher distortion metric but a worse perception metric. Also in Fig. \ref{fig:quantitative results0.3 }, the samples obtained from DVG-Diffusion have more texture detail than DVG-Net. In addition, comparing VPGE-Diffusion and DVG-Diffusion in Table \ref{table:ablation}, the direct use of diffusion model to achieve CT reconstruction can bring good perceptual performance improvement, but it may also lead to contour distortion (as can be seen in Fig. \ref{fig:ablation}). 
% Therefore, to strike a balance between fidelity and perceptual quality, we employ a multi-stage learning approach. This approach utilizes the high-fidelity information $I_{new}$ from $V_{ct}^{init}$ to guide the denoising process of the diffusion model, making the generated model more controllable.
In the task of CT reconstruction from few-view X-ray images, achieving a balance between fidelity and perceptual quality of the reconstructed CT volume remains a significant challenge. This issue is highlighted in \cite{ying2019x2ct}, where two reconstruction approaches were proposed: (1) X2CT-CNN model, which employs a CNN-based method for reconstruction, and (2) X2CT-GAN model, which utilizes a generative GAN-based approach. As shown in Table \ref{table:fid and per}, the X2CT-GAN model achieves better perceptual performance compared to X2CT-CNN (LPIPS decreased from 37.29 to 31.73), but at the cost of a significant reduction in fidelity (SSIM decreased from 0.682 to 0.613). This trade-off is also evident in Fig. \ref{fig:dvg_net_diffusion}, where the structures reconstructed by X2CT-GAN exhibit distortions, while those from X2CT-CNN appear overly smooth.

A similar challenge persists when applying diffusion models. Based on our proposed framework, we constructed three models for comparison: (1) DVG-Net model, which uses a CNN-based method (novel view generation stage) for reconstruction; (2) VPGE-Diffusion model, which employs a diffusion-based approach without additional synthesized views for guidance during our diffusion generation stage; and (3) our complete DVG-Diffusion model. As shown in Table \ref{table:fid and per}, all three models outperform the X2CT series, benefiting from our proposed VPGE, which constructs an aligned latent space that reduces the difficulty of 2D-3D learning. However, purely diffusion-based VPGE-Diffusion model shows better perceptual metrics (LPIPS decreased from 29.50 to 28.11) but worse fidelity (SSIM decreased from 0.702 to 0.667) compared to the CNN-based DVG-Net. In contrast, our DVG-Diffusion model achieves a superior balance between these two aspects, maintaining high fidelity while significantly improving perceptual performance.
This observation is further supported by Fig. \ref{fig:dvg_net_diffusion}, where the VPGE-Diffusion model produces CT volumes with higher visual quality but noticeable structural distortions, while the DVG-Net model reconstructs structures more accurately but suffers from over-smoothed details. Our proposed DVG-Diffusion model effectively combines the strengths of both approaches, delivering superior performance in both image detail and structural integrity. This is attributed to the use of additional new views for guidance in DVG-Diffusion, which are derived from the DVG-Net model's superior structural information. By leveraging these high-quality structural features as guidance, DVG-Diffusion ensures precise CT volume reconstruction without compromising the inherent generative capabilities of the diffusion process.

\begin{table}[h]
  \centering
  \caption{Performance comparison of CT reconstruction between CNN methods and generative methods (GAN,Diffusion).}
  \label{table:fid and per}
% \resizebox{\linewidth}{8mm}{
  \begin{tabular}{c|c|ccc}
    \hline
   
          View&Model&PSNR $\uparrow$  &SSIM $\uparrow$  & LPIPS $\downarrow$  \\
          \hline
 \multirow{5}{*}{Single}& X2CT-CNN \cite{ying2019x2ct} & 22.18 & 0.553  & 44.27\\
 & X2CT-GAN \cite{ying2019x2ct}&  21.40 &  0.505 & 39.62 \\
  &DVG-Net (Ours)&\textbf{23.08}& \textbf{0.568}& 38.59\\
 & VPGE-Diffusion (Ours)& 22.29& 0.532&37.59\\

        &\textbf{DVG-Diffusion (Ours)}&22.54&0.541&\textbf{36.79}\\
        \hline
 \multirow{5}{*}{Biplanar}& X2CT-CNN \cite{ying2019x2ct}
&26.34 &0.682 &37.29\\
 & X2CT-GAN \cite{ying2019x2ct}& 24.82 & 0.613&31.73\\

 & DVG-Net (Ours)& \textbf{27.25}& \textbf{0.702}&29.50\\
 & VPGE-Diffusion (Ours)& 26.47& 0.667&28.11\\
  &\textbf{DVG-Diffusion (Ours)}&26.84&0.679&\textbf{27.35}\\
 \hline

  \end{tabular}
  % }
\end{table}

% \subsection{Learning New View or Learning Other?}
% Our diffusion model is bootstrapped for CT reconstruction by learning new views. Then a new question arises: if we do not learn the new view, can we achieve similar guidance by learning other features? Indeed, we tried direct bootstrapping of the diffusion model conditional on the $Z_{ct}$ (feature level) of the synthesis phase of the new view, but this situation leads to a tendency for the model to converge to a distribution similar to that of $V^{init}_{ct}$, resulting in a reconstructed CT that lacks the necessary perceptual properties (as achieved by the diffusion model). This phenomenon illustrates that the strong prior brought in by the first stage causes the model to converge to some minimal value point, making it difficult to further tune it through the second stage. Therefore, we try to strip out from it the new view $I_{new}$, which is used to compress the features and weaken the distributional prior fitted in the first stage, while still being able to retain reasonable 3D information through the view prior. Also, it is worthy of further investigation whether features can be compressed in other ways to retain valid information in the distribution and achieve higher quality refinement.

\subsection{Clinical Application}
% We utilized real X-ray images obtained from clinical diagnoses in the MIMIC-CXR \cite{johnson2019mimic} dataset for CT reconstruction to validate the robustness of our model on real clinical data. In Fig. \ref{fig:real data}, we provide a visual comparison with X2CT-GAN \cite{ying2019x2ct}. It is evident that our method achieves higher-quality reconstruction performance, producing CT images with more consistent volumetric structures and clearer boundary contours, even in real-world scenarios. We attribute this improvement to the aligned latent space constructed by our method, which bridges the gap between the two modalities. Additionally, the provision of extra views offers supplementary information, eliminating potential spatial inconsistencies in real X-ray images and thereby stabilizing the CT reconstruction process.
To validate the robustness of our model on real clinical data, we utilized authentic X-ray images obtained from clinical diagnoses in the MIMIC-CXR \cite{johnson2019mimic} dataset for CT reconstruction. In Figure \ref{fig:real data}, we present a visual comparison with X2CT-GAN \cite{ying2019x2ct}. Our method demonstrates superior reconstruction quality, generating CT images with more consistent volumetric structures and more well-defined boundary contours, even in real-world scenarios. Furthermore, when reconstructing images with pathological features (e.g., pleural effusion, as shown in the last row), our approach successfully reconstructs clearer 3D pathological features, providing more reliable guidance for surgical planning and prognosis. We attribute these improvements to the aligned latent space constructed by our method, which effectively bridges the gap between the two modalities. Additionally, the incorporation of supplementary views provides complementary information, mitigating potential spatial inconsistencies in real X-ray images and thereby stabilizing the CT reconstruction process.

\subsection{Limitations and Future Work} 
Our DVG-Diffusion model, while effective, has certain limitations. First, the model reconstructs by learning new views to guide the dual-view diffusion model. However, new view extraction takes place in pixel space, while the diffusion process occurs in latent space, necessitating a $z_{ct}$ - $V_{ct}^{init}$ - $I_{new}$ - $z_{new}$ (latent-pixel-latent) process that increases computational burden. Future work aims to streamline this pipeline by operating directly in the latent space, significantly reducing its complexity. 
Furthermore, although our current method achieves state-of-the-art performance, the new view remains subpar compared to the ground truth.
As part of our ongoing efforts, we plan to explore techniques like LLM \cite{achiam2023gpt,song2024pneumollm} or MLLM \cite{li2024llava} to further boost the new view synthesis performance, ultimately leading to improved CT reconstruction results.

\section{Conclusion}
We propose a dual view-guided diffusion model (DVG-Diffusion) to reconstruct CT from few view X-ray images, enhancing reconstruction quality through new-view synthesis and view-guided feature alignment. The proposed view parameter guided encoder (VPGE) facilitates the alignment between 2D X-ray and 3D CT spaces. Additionally, using an auxiliary synthetic X-ray view and input X-ray views to jointly guide the latent diffusion model can reconstruct CT with better perceptual quality. This improvement is achieved through an efficient task decomposition, involving synthesizing new views, translating intricate 2D-to-3D mapping into a more feasible 3D-to-3D latent space mapping, and then reconstructing and refining CT volumes.
{
\bibliographystyle{ieeetr}
\bibliography{x2ct_tip}
}

\end{document}